# Plastic-damage model for concrete in principal directions

S. Ananiev
*Institute of Lightweight Structures and Conceptual Design, University of Stuttgart, Germany*

J. Ožbolt
*Institute of Construction Materials, University of Stuttgart, Germany*

ABSTRACT: In the present paper a plastic-damage model for concrete is discussed. Based on the fact that for isotropic materials the elastic trial stress and the projected plastic stress states have the same eigenvectors, the loading surface is formulated in the principal stress space rather than using the invariants of stress tensor. The model assumes that the directions of orthotropic damage coincide with principal directions of elastic predictor stress state (motivated by coaxial rotated crack model). Due to this assumption, the loading surface and the closest point projection algorithm can still be formulated in the principal directions. The evolution of the inelastic strain is determined using *minimization principle*. Damage and plastic parts of the inelastic strain are separated using a scalar parameter, which is assumed to be stress dependent. The paper also discusses an effective numerical implementation. The performance of the model is demonstrated on one illustrative example.

## 1 INTRODUCTION

Damage and plasticity theories are well established theoretical frameworks for macroscopic (phenomenological) modeling of solids. Application of these theories to the concrete like materials (quasi-brittle materials) suffers from the difficulty in the modeling due to the highly non-symmetrical behavior of such materials under tensile and compressive loading. At the same time, the nature of inelastic deformation is also different. The compressive loading causes the growth of inelastic strains, which are after unloading in a great part irreversible (plastic deformations). In the contrary to this, the tensile loading causes the reduction of stiffness of the material (damage) and the inelastic strains are in a great part reversible.

An elegant way to combine both theories using a single framework was proposed by Meschcke et al. (1998). According to this proposal in the framework of the plasticity theory the inelastic strain can be computed using normality rule. However, only a part of the total inelastic strain is really plastic, i.e. irreversible after unloading. The second part of it is fictive and caused by reduction of stiffness of the material (damage). The separation of these two parts is achieved through a scalar, which is assumed to be a material parameter.

The model developed herein follows this approach, however, the formulation is performed in principal directions which remain constant during closest point projection. This fact was already effectively used in nonlinear plasticity by Simo (1992). An application of this approach in the formulation of a damage model requires an additional assumption – damage-induced anisotropy is reduced to orthotropy and its nature is *"passive"*. This means, if there was a growth of inelastic deformation than the old orientation will be *"erased"* by a new one. The assumption seems to be physically plausible and closely related to the coaxial rotated crack concept. The model represents an alternative approach to the coupling of the rotated crack model with the scalar damage model, as proposed by Jirásek & Zimmermann (1998).

The definition of kinematics of the loading surface (in the case of 2D loading) is done similar to the proposal of Feenstra & de Borst (1996). The surface is parameterized using actual values of tensile and compressive "strengths" of concrete. Their dependencies on the inelastic strains are taken directly from uniaxial experiments. This represents a robust *"mixture"* of kinematic and isotropic hardening with a-priori correct behavior under tension and compression, which is important for practical applications.

The representation of damage through damage variables and the assumption of orthotropy are formally similar to the general framework developed by Carol et al. (2001). The important difference and, at the same time a simplification, is that the evolution laws for damage variables are not formulated directly using normality rules in the space of thermodynamically conjugate forces, which suffer from the lack of physical meaning. The actual values of damage variables are obtained implicitly from the projected of the stress tensor onto loading surface and from the stress equivalence hypothesis. This allows the use of any loading surface and not only that of Rankine ("cut-off") – type.

At the same time, the introduction of damage variables within the framework of orthotropy *"fills"* the tensor product of two normals, used by Meschke et al. (1998) for definition of evolution law of compliance tensor, with physical meaning.

## 2 FORMULATION OF THE MODEL

### 2.1 *Loading surface*

One of the main difficulties in applying theory of plasticity to the concrete-like materials is the proper definition of the shape and kinematic of the loading surface. Under kinematic we understand the evolution of loading surface during increase of inelastic deformations. The most general way to do it offers the use of Haigh-Westergaard coordinates. Several authors have successfully employed this approach for modeling of geomaterials (Willam & Warnke 1975). The disadvantage of this approach is that the models are quite complex what their calibration with the available experimental data make difficult.

For two-dimensional stress-strain conditions, which is the main objective of this paper, the definition of the shape and the kinematic of the loading surface can be considerably simplified. The central idea here was proposed by Feenstra & de Borst (1996). They assumed that the actual shape of the loading surface is for inelastic loading completely described by the actual value of tensile and compressive 'strength', respectively. Their dependencies from accumulated inelastic strains were taken from uniaxial experiments. The obvious and attractive feature of this model is the a-priori correct behavior under uniaxiall tensile and compressive loading. This idea was also exploited in the formulation of the scalar damage model for concrete as presented by Ožbolt & Ananiev (2003).

In the original work of Feenstra & de Borst (1996) the multi-surface plasticity was used. The unique value of the rate of the plastic strain at this point is determined using classical Koiter-rule and active set strategy during closest point projection (see for example Simo & Hughes 1998, Jirásek & Bažant 2002). Although the approach is formally correct, it is physically not correct to assume that an infinitely small rotation of stress tensor causes a finite rotation of plastic strain tensor.

Because of this reason in the present work a single loading surface is formulated based on the experimental evidence (Kupfer et al. 1969) that can be summarized as follows: (i) $f_T$ = uniaxial tensile strength; (ii) $f_C$ = uniaxial compressive strength; (iii) biaxial compressive strength for $\sigma_I=\sigma_{II}$, $f^{1-1}=1.15f_C$; (iv) biaxial compressive strength for $\sigma_I=0.5\sigma_{II}$, $f^{1-0.5}=1.25f_C$ ($f^{1-0.5}$ is the maximal strength). In terms of the here discussed model, the strength (failure envelope) is maximal possible 'strength' of the material. The term actual 'strength' defines the current elastic-plastic-damage state of the material.

One of the most direct ways to construct the loading surface, which satisfies the above experimental data automatically, is to formulate it in polynomial form with respect to eigenvalues (principal stresses) of the stress tensor. The fact that the closest point projection algorithm does not change the orientation of the elastic predictor stress state during projection makes such representation possible. It was already effectively used in nonlinear plasticity by Simo (1992).

After several trials with polynomials of different order, the fourth order polynomial with seven coefficients was chosen. The resulting loading surface reads:

$$f \equiv a_1\left(\sigma_I^2\sigma_{II}^2\right) + a_2\left(\sigma_I^3+\sigma_{II}^3\right) + a_3\left(\sigma_I^2\sigma_{II}+\sigma_I\sigma_{II}^2\right) + \\ + a_4\left(\sigma_I^2+\sigma_{II}^2\right) + a_5\left(\sigma_I\sigma_{II}\right) + a_6\left(\sigma_I+\sigma_{II}\right) + a_7 \quad (1)$$

To determine the seven unknown coefficients ($a_1 – a_7$), seven equations are needed. Four of them can be obtained directly from the experiments. The fifth equation is taken from the fact that for the loading level of $\sigma_I = 0.5\sigma_{II}$ the maximal value of biaxial compressive strength is achieved. The last two equations are obtained from the evidence that localization (crack) in concrete-like material under uniaxial tensile loading is orthogonal to the loading direction. This means that at least for such loading the Rankine "cut-off" criterion is correct. This cri-

terion is assumed to be valid only at a single point. The resulting equation's system reads:

(i) $f(\sigma_I=f_T, \sigma_{II}=0) = 0$

(ii) $f(\sigma_I=f_C, \sigma_{II}=0) = 0$

(iii) $f(\sigma_I=1.15f_C, \sigma_{II}=1.15f_C) = 0$

(iv) $f(\sigma_I=0.5 \cdot 1.25f_C, \sigma_{II}=1.25f_C) = 0$  (2)

(v) $f_{\sigma_{II}}(\sigma_I=0.5 \cdot 1.25 \cdot f_C, \sigma_{II}=1.25 \cdot f_C) = 0$

(vi) $f_{\sigma_I}(\sigma_I=f_T, \sigma_{II}=0) = 1$

(vii) $f_{\sigma_{II}}(\sigma_I=f_T, \sigma_{II}=0) = 0$

Now the equation's system is complete and can be resolved for coefficients of the polynomial. Figure 1 shows the resulting loading surface.

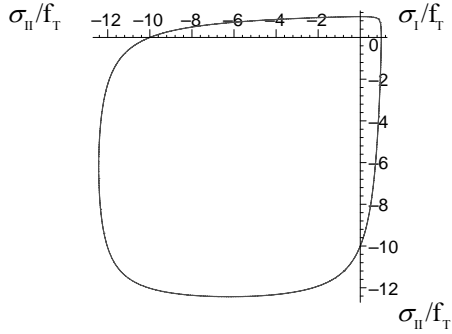

Figure 1. Loading surface as polynomial of order four.

Note that the presented polynomial formulation of the loading surface can be easily extended for 3D loading. To do this, each term in (1) has to be enriched with third direction. For example, the first term $a_1(\sigma_I^2\sigma_{II}^2)$ becomes $a_1(\sigma_I^2\sigma_{II}^2+\sigma_I^2\sigma_{III}^2+\sigma_{II}^2\sigma_{III}^2)$. However, the 3D loading is out scope of the present paper and it will not be discussed here.

2.2 *Kinematics of loading surface*

The presented polynomial form of the loading surface is derived from the failure stage of loading and it represents the failure envelope. It is well known (Jirásek & Bažant 2002) that the inelastic deformations (including damage) start far before the failure is reached. To describe this damage evolution process one has to define the evolution of the shape of loading surface before failure. Because of lack of experimental data here is assumed the same polynomial form of the failure surface.

The dependences of actual strengths of concrete ($f_T$, $f_C$) on the inelastic strains are taken from uniaxial experiments. The central question is how to transfer these dependences to the multiaxial case. Classical plasticity theory distinguishes between isotropic and kinematic hardening. To model isotropic hardening a concept of equivalent strain is introduced. The kinematic hardening is introduced through the concept of back stress. To use similar scheme for concrete, one would require some mixture of kinematic and isotropic hardening. Especially difficult is to find a proper expression for the equivalent strain which would account for the non-symmetrical behavior of concrete under tensile and compressive loading. There are some successful models for concrete, which defines the equivalent strain using only tensile part of strain tensor (Mazars 1986). Here, however, an alternative approach will be employed.

For the 2D case, the kinematic of loading surface can be defined *without* equivalent strain concept. This becomes possible because in the principal stress space the components of total inelastic strain (plastic and damage strain components) can be directly related to the evolution of strengths of concrete. The tensile (positive) part of inelastic strain ($\varepsilon^P$) is used as driving force for evolution of tensile strength and the negative part of inelastic strain is used for evolution of compressive strength, respectively. This idea is illustrated in Figure 2.

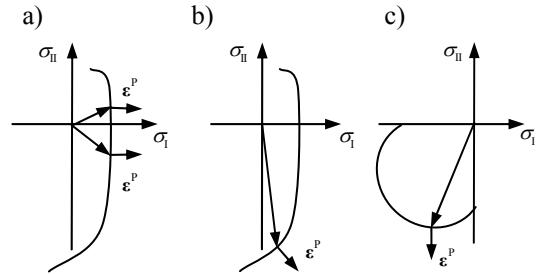

Figure 2. Basic ideas for definition of kinematics of loading surface.

There are three principal cases to be distinguished: (a) Tension-tension or tension-shear loading does *not* affect the compressive strength; (b) Compression-shear loading *does* affect the both strengths; (c) Compression-compression loading does *not* affect the tensile strength. The above definition for evolution of concrete strength fulfills two important requirements of the macroscopic concrete model: (i) the uniaxial tensile loading does not affect the strength in orthogonal direction and (ii) the failure under compression means a full loss of load bearing capacity in orthogonal direction.

## 2.3 Extension to damage

The formulation of the presented model is performed in the principal axes which do not rotate during closest point projection algorithm. This, actually, predefines the structure of damage which can be modeled within the present approach. Damage can be treated as a scalar damage, as already presented by Ožbolt & Ananiev (2003), or it can be considered as an orthotropic – *"passive"* damage, which will be presented here. The term *"passive"* means that during inelastic loading the old orientation of two damage directions will be *"erased"* by the new one.

### 2.3.1 Damage variables

In 2D case the orthotropic material can be described by five elastic parameters (Herakovich 1998): (i) Young's modulus in both principal directions: $E_1, E_2$; (ii) Poisson's ratios in orthogonal directions: $\mu_{12}, \mu_{21}$; (iii) shear modulus $G$.

By requiring of symmetry for the elastic modulus, one obtains the reciprocal relation which reduces the number of independent material parameters to four. The ratio between Young's modulus and between Poisson's ratios has to be the same:

$$n = \frac{E_1}{E_2} = \frac{\mu_{12}}{\mu_{21}} \qquad (3)$$

This characterizes the "degree of orthotropy" in the material ($n$). Taking into account (3) the generalized elasticity law for 2D case reads:

$$\begin{bmatrix} \sigma_x \\ \sigma_y \\ \tau \end{bmatrix} = \frac{E_2}{1-n\mu_{21}^2} \begin{bmatrix} n & n\mu_{21} & 0 \\ n\mu_{21} & 1 & 0 \\ 0 & 0 & \overline{G} \end{bmatrix} \begin{bmatrix} \varepsilon_x \\ \varepsilon_y \\ \gamma \end{bmatrix} \qquad (4)$$

Following the classical scalar damage formulation, as introduced by Kachanov (1958), for orthotropy two damage variables $d_1$ and $d_2$ are introduced:

$$\begin{aligned} E_1 &= (1-d_1)E_0, & \mu_{12} &= (1-d_1)\mu_0 \\ E_2 &= (1-d_2)E_0, & \mu_{21} &= (1-d_2)\mu_0 \end{aligned} \qquad (5)$$

To keep the symmetry of the elasticity matrix, in (4) the same relations should be valid for Young's modulus and for Poisson's ratios. From engineering point of view this requirement is quite reasonable because in case of fully localized damage two directions of orthotropy are totally uncoupled, as illustrated in Figure 3.

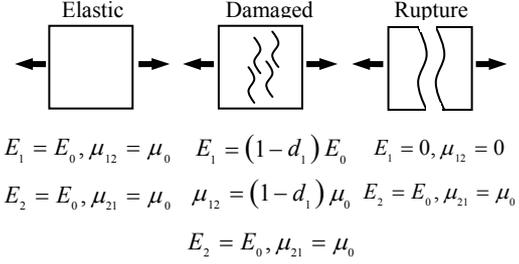

$$E_1 = E_0, \mu_{12} = \mu_0 \quad E_1 = (1-d_1)E_0 \quad E_1 = 0, \mu_{12} = 0$$
$$E_2 = E_0, \mu_{21} = \mu_0 \quad \mu_{12} = (1-d_1)\mu_0 \quad E_2 = E_0, \mu_{21} = \mu_0$$
$$E_2 = E_0, \mu_{21} = \mu_0$$

Figure 3. Definition of the orthotropic damage variables.

It remains to connect the shear modulus with introduced damaged variables. To establish the required relation, the idea of Darwin & Pecknold (1977) is employed. They assumed that the shear modulus of cracked concrete can be determined from the condition of its invariance with respect to the rotation of coordinate system. This idea is illustrated by the following equations ($\theta$ is the angle between the coordinate systems x-y and a-b):

$$\begin{aligned} \boldsymbol{\sigma}_{a\text{-}b} &= \mathbf{D}_{a\text{-}b} : \boldsymbol{\varepsilon}_{a\text{-}b} \\ \boldsymbol{\varepsilon}_{a\text{-}b} &= \mathbf{T}_\varepsilon \boldsymbol{\varepsilon}_{x\text{-}y}, \boldsymbol{\sigma}_{a\text{-}b} = \mathbf{T}_\sigma \boldsymbol{\sigma}_{x\text{-}y} \\ \mathbf{D}_{x\text{-}y} &= \mathbf{T}_\sigma^{-1} \mathbf{D}_{a\text{-}b} \mathbf{T}_\varepsilon \\ \overline{G}_{x\text{-}y} &\equiv \left(\cos^2\theta - \cos^4\theta\right)\cdot\left[1 - 2n\mu_{21} + n - 4\overline{G}_{a\text{-}b}\right] + \overline{G}_{a\text{-}b} \end{aligned} \qquad (6)$$

Condition of invariance requires:

$$1 - 2n\mu_{21} + n - 4\overline{G}_{a\text{-}b} \equiv 0 \qquad (7)$$

The last equation leads to the following final expression for shear modulus:

$$G_{a\text{-}b} = \frac{E_2}{1-n\mu_{21}^2}\left[\frac{1}{4} - \frac{1}{2}n\mu_{21} + \frac{1}{4}n\right] \qquad (8)$$

Introducing (5) into (8) one can express shear modulus $G_{a\text{-}b}$ as a function of damage. This eliminates the necessity in *shear retention factors*, which are widely used in rotated crack models.

It is important to point out that despite the invariance of $G$ the orthotropy of the material is preserved. This means that the usual classification of anisotropic materials is not complete (see Herakovich 1998). There is a material type which lies between transversely isotropic and fully orthotropic material.

### 2.3.2 Evolution laws

Within the framework of the presented model, the inelastic stains can be of different nature, i.e. plastic or damage strains. If they are resulting from degradation of the material stiffness than they are fictitious and the actual quantities which grow are damage variables. The key idea in the definition of the evolution laws is the fact that during inelastic loading it is *not possible* to find out what kind of inelasticity took place. This will be clear only upon unloading. This allows adaptation of the classical scheme used in plasticity, where the evolution laws are defined for inelastic strains and not for damage variables. The obvious advantage is the possibility for the use of the loading surface that is formulated in the stress space, which possesses a clear physical meaning.

The above idea was already used by Meschke et al. (1998). They defined an evolution law directly for compliance tensor **C** using tensor product of two normals:

$$\dot{\mathbf{C}} = \dot{\gamma} \frac{\partial_\sigma f \otimes \partial_\sigma f}{\partial_\sigma f : \boldsymbol{\sigma}} \quad (9)$$

Formally, this expression is correct, i.e. product of two second order tensors gives tensor of order four. However, from the engineering point of view it would be interesting to clarify what kind of anisotropy is induced by damage growth. Moreover, it would be useful to know a direct relation between damage variables and coefficients of compliance tensor. These questions are discussed for instance by Carol et al. (2001), where the second order damage tensor was constructed and an explicit relation between it and the secant material stiffness was derived using *energy equivalence principle* and *product-type* symmetrization. The stiffness of an orthotropic material can be easily represented by the secant formulation given by Carol et al. (2001). For instance, in 2D case, five elastic parameters from (5) are related to the principal values of damage tensor as:

$$E_1 = (1-d_1)^2 E_0, \quad \mu_{12} = \frac{(1-d_2)}{(1-d_1)} \mu_0$$

$$E_2 = (1-d_2)^2 E_0, \quad \mu_{21} = \frac{(1-d_1)}{(1-d_2)} \mu_0 \quad (10)$$

$$G_{12} = (1-d_1)(1-d_2) \frac{1}{2} \frac{E_0}{(1+\mu_0)}$$

These relations do satisfy the basic requirement of orthotropy (3), however, for the case of isotropic damage ($d_1=d_2$) they yield to unrealistic prediction: during increase of damage Poisson's ratio remains constant. It seems physical to assume that in case of complete isotropic rupture *all* elastic constants yield to zero. The expression for shear modulus also suffers from the lack of physical meaning. In the case of fully localized rupture in one particular direction (see Figure 3) the shear stiffness is definitely not zero. For example, (8) yields for this case to the shear modulus of $0.25E_2$. The *sum-type* symmetrization fulfills the last requirement, but still fails at first one, because it results in the same relations for Poisson's ratio as the product-type symmetrization.

The real difficulties begin by the formulation of the evolution laws. Applying the formal structure of plasticity, one has to define a flow rule for damage variables in the space of thermodynamically conjugate forces. Unfortunately, these forces do not have physical meaning. Consequently, it is difficult to formulate a meaningful loading surface which is the main ingredient of the formulation. In Carol et al. (2001) this problem is partially solved through introduction of *pseudo-logarithmic* damage tensor. But their formulation is restricted to Rankine-type of loading surface. From engineering point of view, this is a severe restriction, because material can fail also in compression (crushing of concrete).

Here, the advantages of both approaches are combined, i.e. the evolution law of damage variables is defined using loading surface in stress space. A standard way to introduce the associated flow rule is the maximum dissipation postulate. It states that among all possible stresses only those will be real which impose maximum to the dissipation function under condition that the stress state does not violate the loading surface. Its numerical integration within the framework of backward Euler scheme results in closest point projection of elastic stress state onto the loading surface (see Simo & Hughes 1998). This projection can be formulated as optimization problem, where unknown are the elastic strains ($\boldsymbol{\varepsilon}^{\text{Elastic}}$) at the (n+1) load increment ($\mathbf{D}_0$ = elastic stiffness tensor):

$$\min_{f\left(\boldsymbol{\varepsilon}^{\text{Elastic}}_{n+1}\right)\leq 0} \frac{1}{2} \left(\boldsymbol{\varepsilon}^{\text{Total}}_{n+1} - \boldsymbol{\varepsilon}^{\text{Elastic}}_{n+1}\right) : \mathbf{D}_0 : \left(\boldsymbol{\varepsilon}^{\text{Total}}_{n+1} - \boldsymbol{\varepsilon}^{\text{Elastic}}_{n+1}\right) \quad (11)$$

Recently, several authors (Ortiz & Repetto 1999, Miehe 2003) have generalized this formulation to the *incremental minimization principles* for standard dissipative materials.

If the loading surface is represented by the polynomial of order higher than two, the solution of above stated optimization problem is not an easy task, and will be discussed in the next section. Let's assume that the actual values of elastic strain are found. The *stress equivalence hypothesis* for the eigenvalues of converged stress state results in the two equations which are sufficient for determination of both orthotropic damage variables ($d_1, d_2$). With other words, when restricting ourselves to orthotropic damage there is *no need* for any evolution laws of damage variables in explicit form. They are obtained by solving the following equation system:

$$\mathbf{D}_{n+1} : \left( \boldsymbol{\varepsilon}_{n+1}^{Total} - \beta(\alpha_{n+1}) \cdot \gamma_{n+1} \frac{\partial f}{\partial \boldsymbol{\sigma}_{n+1}} \right) \equiv$$
$$\equiv \mathbf{D}_0 : \left( \boldsymbol{\varepsilon}_{n+1}^{Total} - \gamma_{n+1} \frac{\partial f}{\partial \boldsymbol{\sigma}_{n+1}} \right) \quad (12)$$

where $\mathbf{D}_{n+1}$ follows from equations (3-8) and the total inelastic strain including Lagrange multiplier $\gamma_{n+1}$, is known from (11). The parameter $\beta$ has the same purpose as in Meschke et al. (1998). It separates the damage part of inelastic strain from its plastic part. In the original work, this parameter is assumed to be the material constant. However, it seems more physical to assume it to be stress depended. The function $\beta(\alpha)$ can be, for example, linear. The parameter $\alpha$ is defined as a relation between the total stored elastic energy and the work done by the compressive stresses:

$$\alpha = \frac{\langle \sigma_I \rangle_-^2 - 2\mu_0 \langle \sigma_I \rangle_- \langle \sigma_{II} \rangle_- + \langle \sigma_{II} \rangle_-^2}{\sigma_I^2 - 2\mu_0 \sigma_I \sigma_{II} + \sigma_{II}^2} \quad (13)$$

where $\langle \sigma \rangle_- = 0$ if $\sigma > 0$ and $\langle \sigma \rangle_- = \sigma$ if $\sigma < 0$. The important feature of the parameter $\alpha$ is that it is always positive and bounded between zero and one. Value 'one' means full compressive loading with maximal contribution of the plastic part to the total inelastic strain. Here, the advantage of the formulation in the principal directions is utilized again.

## 3 INTEGRATION ALGORITHM

The presented polynomial form of the loading surface allows a relatively good approximation of the experimental data for concrete. However, this form has a serious drawback – the existence of artificial elastic regions. The reason for this is the order of the polynomial formulation. If it is higher than two, than in general case the polynomial has a more than one extreme, what means that the loading surface can be negative in more than one region. To assure a robust convergence of the standard closest point projection algorithm the load increments should be rather small. In our case the enrichment of the objective function with penalty term, as recently proposed by Armero & Pérez-Foguet (2002) is not helpful. Namely, if the elastic predictor stress state lies in artificial elastic region than the projection algorithm will simply not start. Therefore, an alternative approach is developed which is based on *"methods of centers"*, that was originally proposed by Bui-Trong-Lieu & Huard (1966) (see also Polak 1971). However, before we proceed with the formulation of the method, an alternative formulation of Karush-Kuhn-Tucker condition is discussed that is helpful for understanding of the proposed approach.

### 3.1 *Alternative formulation of Karush-Kuhn-Tucker condition*

The classical definition of optimality condition of constrained optimization problem (see Luenberger 1989) states the existence of Lagrangian multipliers only in optimum, where the gradient of objective function can be expressed as a linear combination of gradients of constraints active at this point. Such linear dependence means that there is no further possibility to reduce the objective function without violating of active constraints.

The main problem of this formulation is its implicit character. It states only the existence of multipliers, but does not say how to find them. Probably because of this reason, the usual way to solve this optimality condition is to write down the complete equation system and to apply the Newton's method. In optimization theory, this approach is called Sequential Quadratic Programming (SQP). The term 'quadratic' comes from the fact that the application of Newton's method to KKT equation system means that in the each iteration the original objective function is replaced by its quadratic approximation. The algorithm has a nice property of *local* quadratic rate of convergence. However, if the initial approximation lies far from optimum than the algorithm may not converge, which is a well known problem of Newton's method.

On the other hand, it can be shown that the Lagrangian multipliers exist not only in optimum, but also at the any point belonging to active constraint. Such definition was proposed for the first time by Hestenes (1975) (see also Rockafellar 1993). According to his proposal, the multipliers are introduced not at the optimum, but *during projection* of objective function's gradient onto the tangential space of constraints. The projection is defined as a linear combination in the following way:

$$\vec{\mathbf{p}} = \nabla \vec{\mathbf{f}} + \sum_{i}^{NC} \gamma_i \nabla \vec{\mathbf{h}} \quad \& \quad \vec{\mathbf{p}} \cdot \nabla \vec{\mathbf{h}}_i = 0 \Rightarrow \vec{\mathbf{p}} = 0 \quad (14)$$

where, ∇**f** - objective's function gradient, ∇**h** - gradient of active constraints, NC - number of active constraints, $\chi_i$ - "generalized" Lagrangian multipliers. The actual values of 'generalized' multipliers are found from the condition of orthogonality between projected gradient and *each gradient* of active constraints. Under requirement of linear independence of vectors, the resulting linear equation's system allows a unique definition of multipliers. If the resulting projected gradient is a zero vector, than a local optimum is obtained and the multipliers get their classical meaning. A simple proof of the sufficiency of this optimality condition can be found in Ananiev (2003).

### 3.2 *Method of centers*

As already mentioned, the presence of artificial elastic regions in the loading surface requires a reconsideration of standard closest point projection algorithm. If the elastic predictor lies in such elastic region, than the algorithm will not start. Even if the elastic predictor does not lie in such areas, the Newton's method may converge to the wrong surface, because of the negative curvature at the starting point.

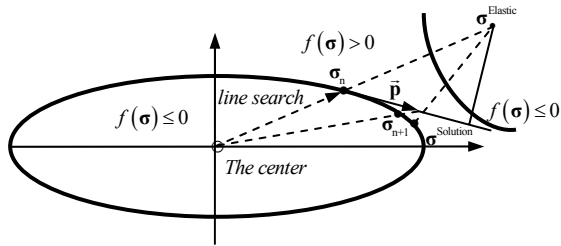

Figure 4. One iteration in the method of centers.

To guarantee the convergence to the right surface the integration algorithm has to generate a sequence of stress states which do not leave this surface. One possibility to achieve this goal offers the *"method of centers"*, developed by Bui-Trong-Lieu & Huard (1966). The method introduces some base point through which the "correct" surface is marked. At the each iteration step the projected gradient (14) is generated and than the stress state is changed according to this direction. If the loading surface is convex, the resulting stress state lies outside of the loading surface. The corrected value, which lies exactly at the surface, is obtained through line search along the line connecting the current estimate and the center, which lies inside the surface. From all possible solutions the minimal value is chosen. The algorithm is considered as converged if the norm of projected gradient is zero. The iteration scheme of the algorithm is illustrated in Figure 4.

## 4 NUMERICAL EXAMPLE

### 4.1 *Nooru-Mohamed test*

The experimental investigations of mixed-mode failure in plain concrete was carried out by Nooru-Mohamed (1992). The main difficulty in this test is the continuous rotation of principal directions of the stress state. If the model is not capable to represent a stress-induced anisotropy, than this can be immediately seen on the resulting crack pattern. For instance, a scalar damage model proposed in Ožbolt & Ananiev (2003) showed a correct behavior at the structural level but failed to produce the two separate cracks, as observed in the experiment. Here, we use again this experiment ($P_H$ = 10kN) to test the presented model. It should be noted that the test allows to examine only those part of the proposed formulation which concerns the kinematics of loading surface. With this test it is not possible to check the implicit evolution laws for damage, because there was no unloading at the structural level. For concrete, a simple bi-linear tensile stress-strain law was used. The calculated and in the experiment measured load-displacement curves are shown in Figure 5. Figure 6a shows calculated crack pattern (cracks = darks zones – maximal principal strains) and Figure 6b shows experimentally obtained cracks. As can be seen, the calculated results agree well with the experimental observations. The model is able to predict the correct crack path what means that the damage induced anisotropy is accounted for correctly.

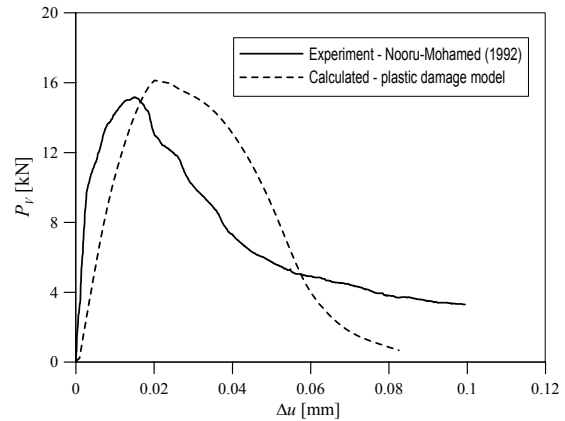

Figure 5. Load-displacement diagram.

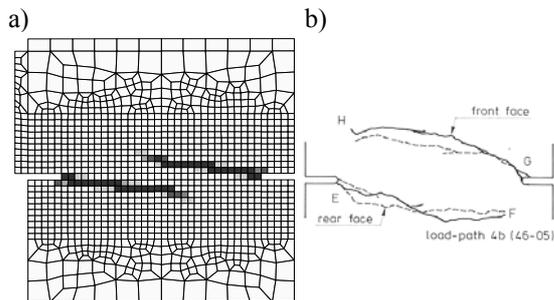

Figure 6. Calculated and experimental crack patterns.

## 5 CONCLUSIONS

In the presented work, the authors have attempted to reconsider some basic concepts, which are widely used in computational modeling of inelastic behavior of solid materials. Rejection of using the *equivalent strain*, *thermodynamically conjugate forces* and *Karush-Kuhn-Tucker condition* allowed not only to formulate the model with clear and intuitive structure of relations between internal variables, but it allows also the development of an effective numerical algorithm. Application of this model to relatively complex mixed-mode fracture of plain concrete can serve as a justification of proposed approach. Nevertheless, it is also obvious that more work is needed to refine the concept and to apply it for the modeling of other types of damage phenomena in concrete.

## 6 REFERENCES

Ananiev, S. 2003. An alternative solution of optimal material distribution problem. In: Bestle D. et al (eds.). *EUROMECH 442 Colloquium on Computer-Aided Optimization of Mechanical Systems*. Erlangen-Nuremberg, Germany.
Armero, F., Perez-Foguet, A. 2002. On the formulation of closest-point projection algorithms in elastoplasticity—part I: The variational structure. *International Journal for Numerical Methods in Engineering*, 53: 297–329.
Bui-Trong-Lieu, Huard, P. 1966. La méthode des centers dans un espace topologique. *Numerische Mathematik*, 8: 65-67.
Carol, I., Rizzi, E., Willam, K. 2001. On the formulation of anisotropic elastic degradation. I. Theory based on a pseudo-logarithmic damage tensor rate. *International Journal of Solids and Structures*, 38: 491-518.
Darwin, D., Pecknold D. 1977. Analysis of plane R/C structures. *Computers & Structures*, 7: 137-147.
Feenstra, P., de Borst, R. 1996. A composite plasticity model for concrete. *International Journal of Solids and Structures*, 33: 707-730.
Herakovich, C. 1998. *Mechanics of Fibrous Composites*. John Willey & Sons, New York.
Hestenes, M.R. 1975. *Optimization theory. The finite dimensional case*. John Wiley & Sons, New York.
Jirásek, M., Bažant, Z. 2002. *Inelastic Analysis of Structures*. John Willey & Sons, West Sussex.
Jirásek, M., Zimmermann, T. 1998. Rotating crack model with transition to scalar damage. *Journal of Engineering Mechanics ASCE*, 124: 277-284.
Kachanov, L.M 1958. Time of the rupture process under creep conditions. *Izvestija AN SSSR, Otdel Technicheskih Nauk*, 8: 26-31. (in Russian)
Kupfer, H., Hilsdorf, H.K., Rüsch, H. 1969. Behavior of concrete under biaxial stress. *ACI Journal*, 656-666.
Luenberger, D. 1989. *Linear and nonlinear programming*. Addison-Wesley, Second edition.
Mazars, J. 1986. A description of micro- and macro-scale damage of concrete structures. *Journal of Engineering Fracture Mechanics*, 25: 729-737.
Meschke, G., Lackner, R., Mang, H. 1998. An anisotropic elastoplastic-damage model for plain concrete. *International Journal of Numerical Methods in Engineering*, 42: 703-727.
Miehe, C. 2003. Microstructure development in standard dissipative solids based on incremental energy minimization principles. In: Oñate E. et al. (eds.). *COMPLAS 2003 VII International Conference on Computational Plasticity*. CIMNE, Barcelona.
Nooru-Mohamed, M.B. 1992. *Mixed-mode fracture of concrete: an experimental approach*. Ph.D. Thesis, TU Delft.
Ortiz, M., Repetto, E.A. 1999. Nonconvex energy minimization and dislocation structure in ductile single crystals. *Journal of the Mechanics and Physics of Solids*, 47: 397–462.
Ožbolt, J., Ananiev, S. 2003. Scalar damage model for concrete without explicit evolution law. In: Bićanić, N., et al (eds.). *Proceedings of the EURO-C 2003 Conference on Computational Modelling of Concrete Structures*. Swets & Zeitlinger B.V., Lisse.
Polak, E. 1971. *Computational methods in optimization. A unified approach*. Academic Press, New York.
Rockfellar, R.T. 1993. Lagrange multipliers and optimality. *SIAM Review*, 35(2): 183-238.
Simo, J. 1992. Algorithms for static and dynamic multiplicative plasticity that preserve the classical return mapping schemes of the infinitesimal theory. *Computer Methods in Applied Mechanics and Engineering*, 99: 61-112.
Simo, J.C., Hughes, T.J.R. 1998. *Computational inelasticity*. Springer, New York.
Willam, K., Warnke, E. 1975. Constitutive models for the triaxial behavior of concrete. In: *Seminar on "Concrete Structures Subjected to Triaxial Stresses"*, International Association of Bridge and Structural Engineering (IABSE), Bergamo, Italy, 19: 1-30.

## ACKNOWLEDGEMENTS

A financial support for the first author from the German Academic Exchange Service (DAAD) is gratefully acknowledged.

## APPENDIX

The frame invariance of proposed formulation is guaranteed due to the fact that the growth of damage variables takes place only if elastic predictor lies out of the loading surface. The closest point projection is performed exactly as in plasticity, where the main unknowns are inelastic strains and not damage variables (see Eq.11 & 12).